\documentclass[10pt,a4papper]{article}
\usepackage[utf8]{inputenc}
\usepackage[english]{babel}
\usepackage{indentfirst}
\usepackage[]{graphicx}
\usepackage{geometry} 
\begin{document}
	\title{\textbf{The decay $\tau \to K^{*0}(892) K^- \nu_{\tau}$ in the extended NJL model}}
	\author{M. K. Volkov$^{1}$\footnote{volkov@theor.jinr.ru}, A. A. Pivovarov$^{1}$\footnote{tex$\_$k@mail.ru}, K. Nurlan$^{1 2 3}$\footnote{nurlan@theor.jinr.ru}\\
		\small
		\emph{$^{1}$Joint Institute for Nuclear Research, Dubna, 141980, Russia}\\	
		\small
		\emph{$^{2}$Institute of Nuclear Physics, Almaty, 050032, Kazakhstan}\\
		\small
		\emph{$^{3}$L. N. Gumilyov Eurasian National University, Nur-Sultan, 01008, Kazakhstan}}
	\date{}
	\maketitle
	\small
	
\begin{abstract}
In the extended Nambu -- Jona-Lasinio model, the decay width of $\tau \to K^{*0}(892) K^- \nu_{\tau}$ is calculated. The contributions from the contact diagram and diagrams with intermediate axial-vector, vector and pseudoscalar mesons in the ground and first radially excited states are taken into account. It is shown that axial-vector and vector channel with a contact diagram give a dominant contribution to the decay width. The obtained results for the decay width $\tau \to K^{*0}(892) K^- \nu_{\tau}$ are in satisfactory agreement with experimental data. A prediction for differential distribution over the invariant mass of the meson pair $K^{*0}(892) K^-$ is given. 
\end{abstract}

\large
\section{Introduction}
	Decays of the $\tau$ lepton into hadrons offer a unique possibility to study the electroweak charged hadronic current at the energy scale not exceeding the mass value of the $\tau$. The study of $\tau$ decays with strange particles is especially interesting, because all possible intermediate channels work in it at once: contact channel, when the final states are directly produced from lepton current without any intermediate meson states, channels with axial-vector, vector, and pseudoscalar mesons in the ground and first radially excited states. Studying these channels allows one to check the QCD phenomena at low-energies, the violation of $U(3)\times U(3)$ symmetry and the fulfillment of predictions of the vector dominance model.
	
	The process $\tau \to K^{*0}(892) K^- \nu_{\tau}$ was repeatedly investigated from both the experimental \cite{Albrecht:1995wx, Barate:1997ma, Tanabashi:2018oca} and theoretical \cite{Li:1996md, Guo:2008sh, Dai:2018thd} points of view. The data for these studies were collected with the ARGUS detector at the storage ring DORIS II of DESY \cite{Albrecht:1995wx} and with the ALEPH detector at the LEP \cite{Barate:1997ma}. 
	
	Theoretical calculations of the process $\tau \to K^{*0}(892) K^- \nu_{\tau}$ were carried out in the framework of the $U(3) \times U(3)$ chiral-symmetric model \cite{Li:1996md} and chiral theories with resonances \cite{Guo:2008sh}. Also, in the recent theoretical paper \cite{Dai:2018thd}, the angular momentum algebra was applied.

	In the present paper, in the framework of the extended Nambu -- Jona-Lasinio (NJL) model \cite{Volkov:1996br, Volkov:1996fk, Volkov:2005kw, Volkov:2017arr}, the decay width of $\tau \to K^{*0}(892) K^- \nu_{\tau}$ is calculated and the differential distribution over the invariant mass of the meson pair $K^{*0}(892) K^-$ is predicted. 
	
	The extended NJL model allows one to describe mesons not only in the ground states but also in the first radially excited meson states and their interactions at low energies without violating $U(3) \times U(3)$ chiral symmetry. In recent years,  by using the extended NJL model, the meson production processes in electron-positron annihilation and many of main $\tau$ lepton decays are described without any additional free parameters \cite{Volkov:2017arr}.
	
\section{Quark-meson Lagrangian of the extended NJL model}
In the extended NJL model, part of the quark–meson interaction Lagrangian referring to the mesons involved in the process under consideration has the form \cite{Volkov:1996fk, Volkov:2017arr}:

	\begin{eqnarray}
	\label{Lagrangian}
		\Delta L_{int} & = &
		\bar{q} \biggl[ \frac{1}{2} \gamma^{\mu} \sum_{j=\pm, 0} \lambda_{j}^{K} \left(A_{K^{*}}K^{*j}_{\mu}  + B_{K^{*}}K^{*'j}_{\mu}\right) + i \gamma^{5} \sum_{j = \pm} \lambda_{j}^{K} \left(A_{K}K^{j} + B_{K}K'^{j}\right) \nonumber \\ 
		&& +\frac{1}{2} \gamma^{\mu} \sum_{j = \pm} \lambda_{j}^{\rho} \left(A_{\rho}\rho^{j}_{\mu} + B_{\rho}\rho'^{j}_{\mu} \right)	+ \frac{1}{2} \gamma^{\mu} \gamma^{5} \sum_{j=\pm} \lambda_{j}^{\rho} \left(A_{a_1}{a_1}^{j}_{\mu} + B_{a_1}{a'_1}^{j}_{\mu}\right) \nonumber \\ 
		&& +  i \gamma^{5} \sum_{j = \pm} \lambda_{j}^{\pi} \left(A_{\pi}{\pi}^{j} + B_{\pi}{\pi'}^{j}\right) \biggl]q,
	\end{eqnarray}
where, $q$ and $\bar{q}$ are the u, d, and s quark fields with the constituent masses $m_{u} \approx m_{d} = 280$~MeV, $m_{s} = 420$~MeV; the prime marks excited meson states and $\lambda$ are linear combinations of the Gell-Mann matrices \cite{Volkov:2017arr},
	\begin{eqnarray}
	\label{verteces}
	\label{coupling}
		A_{M} & = & \frac{1}{\sin(2\theta_{M}^{0})}\biggl[g_{M}\sin(\theta_{M} + \theta_{M}^{0}) + g'_{M}f_{M}(k_{\perp}^{2})\sin(\theta_{M} - \theta_{M}^{0})\biggl], \nonumber\\
		B_{M} & = & \frac{-1}{\sin(2\theta_{M}^{0})}\biggl[g_{M}\cos(\theta_{M} + \theta_{M}^{0}) + g'_{M}f_{M}(k_{\perp}^{2})\cos(\theta_{M} - \theta_{M}^{0})\biggl].
	\end{eqnarray}
The subscript $M$ specifies the corresponding meson.  
	
	The formfactor $f\left(k_{\perp}^{2}\right) = \left(1 + d k_{\perp}^{2}\right)\Theta(\Lambda^{2} - k_{\perp}^2)$ describes the first radially excited meson states,  $\Lambda = 1.03$ GeV is the cutoff parameter. The slope parameter $d$ is depends only on the quark composition of the corresponding meson \cite{Volkov:1996fk, Volkov:2017arr}:
	\begin{eqnarray}
		d_{uu} = -1.784 \times 10^{-6} \textrm{MeV}^{-2}, \nonumber \\
		d_{us} = -1.761 \times 10^{-6}\textrm{MeV}^{-2}.
	\end{eqnarray}
	
	This parameter is unambiguously fixed from the condition of constancy of the quark condensate after the inclusion of radially excited states. 
	The relative transverse momentum of the internal quark–antiquark pair is:
	\begin{eqnarray}
		k_{\perp} = k - \frac{(kp) p}{p^2},
	\end{eqnarray}
	where $p$ is the meson momentum. In the meson rest frame,
	\begin{eqnarray}
		k_{\perp} = (0, {\bf k}).
	\end{eqnarray}
	
Therefore, this momentum can be used in the three-dimensional form.

The parameters $\theta_{M}$ are the mixing angles determined after diagonalizing the free Lagrangian for the ground
and first radially excited states \cite{Volkov:1996fk,Volkov:2017arr}:
	\begin{eqnarray}
		\theta_{K^{*}} = 84.74^{\circ}, &\quad& \theta_{K} = 58.11^{\circ}, \nonumber\\
		\theta_{\rho} = \theta_{a_1} = 81.8^{\circ}, &\quad& \theta_{\pi} = 59.48^{\circ}.
	\end{eqnarray}
	
In addition, $\theta_{M}^{0}$ are auxiliary parameters introduced for convenience as
	\begin{eqnarray}
	\label{tetta0}
		&\sin\left(\theta_{M}^{0}\right) = \sqrt{\frac{1 + R_{M}}{2}},& \nonumber\\
		& R_{K^{*}} = \frac{I_{11}^{f_{us}}}{\sqrt{I_{11}I_{11}^{f^{2}_{us}}}}, \quad R_{K} = \frac{I_{11}^{f_{us}}}{\sqrt{Z_{K}I_{11}I_{11}^{f^{2}_{us}}}},& \nonumber\\
		&R_{\rho} = R_{a_1} = \frac{I_{20}^{f_{uu}}}{\sqrt{I_{20}I_{20}^{f^{2}_{uu}}}}, \quad R_{\pi} = \frac{I_{20}^{f_{uu}}}{\sqrt{Z_{\pi} I_{20}I_{20}^{f^{2}_{uu}}}},&
	\end{eqnarray}
	where
	\begin{eqnarray}
	\label{Zpi}
		Z_{\pi} = \left[1 - \frac{6m^2_{u}}{M^2_{a_1}} \right]^{-1},
	\end{eqnarray}
	\begin{eqnarray}
	\label{Zk}
		Z_{K} = \left[1 - \frac{3}{2}(m_{u} + m_{s})^{2}\left(\frac{\sin^{2}{\alpha}}{M^{2}_{K_{1(1270)}}} + \frac{\cos^{2}{\alpha}}{M^{2}_{K_{1(1400)}}}\right)\right]^{-1}.
	\end{eqnarray}
	
	Here $Z_{\pi}$ and $Z_K$ are the additional renormalization constants appearing in the $\pi$--$a_1$ and $K$--$K_1$ transitions. $M_{a_{1}(1260)} = 1299 (+12, -28)$~MeV, $M_{K_{1}(1270)} = 1272 \pm 7$~MeV, $M_{K_{1}(1400)} = 1403 \pm 7$~MeV are the mass of axial-vector mesons \cite{Tanabashi:2018oca, Akhunzyanov:2018lqa}. The splitting of
the $K_{1A}$ state into two physical mesons $K_{1}(1270)$ and $K_{1}(1400)$ is taken into account in the expression for $Z_K$, $\alpha=57^{\circ}$ \cite{Volkov:2019yhy, Volkov:2019cja}. 
	
	Integrals appearing in quark loops are 
	\begin{eqnarray}
	I^{f^m}_{n_{1}n_{2}} =
	-i\frac{N_{c}}{(2\pi)^{4}}\int\frac{ f^m({\bf k}^2)}{(k^2 - m_{u}^{2}  )^{n_{1}}(k^2 - m_{s}^{2} )^{n_{2}}}
	\mathrm{d}^{4}k.
	\end{eqnarray}
	
	The quark-meson coupling constants have the form
	\begin{eqnarray}
	\label{Couplings}
	g_{K^{*}} = \left(\frac{2}{3}I_{11}\right)^{-1/2}, &\quad& g_{K} =  \left(\frac{4}{Z_{K}}I_{11}\right)^{-1/2},  \nonumber\\
	g_{K^{*'}} = \left(\frac{2}{3}I^{f^2}_{11}\right)^{-1/2}, &\quad& g_{K'} =  \left(4I^{f^2}_{11}\right)^{-1/2},  \nonumber\\
	g_{\rho} =  \left(\frac{2}{3}I_{20}\right)^{-1/2}, &\quad& g_{\pi} =  \left(\frac{4}{Z_{\pi}}I_{20}\right)^{-1/2}, \nonumber\\
	g_{\rho'} =  \left(\frac{2}{3}I^{f^2}_{20}\right)^{-1/2}, &\quad& g_{\pi'} =  \left(4I^{f^2}_{20}\right)^{-1/2}, .
	\end{eqnarray}	
	
\section{The process $\tau \to K^{*0}(892) K^- \nu_{\tau}$ in the extended NJL model}
\begin{figure}[h]
		\center{\includegraphics[scale = 0.8]{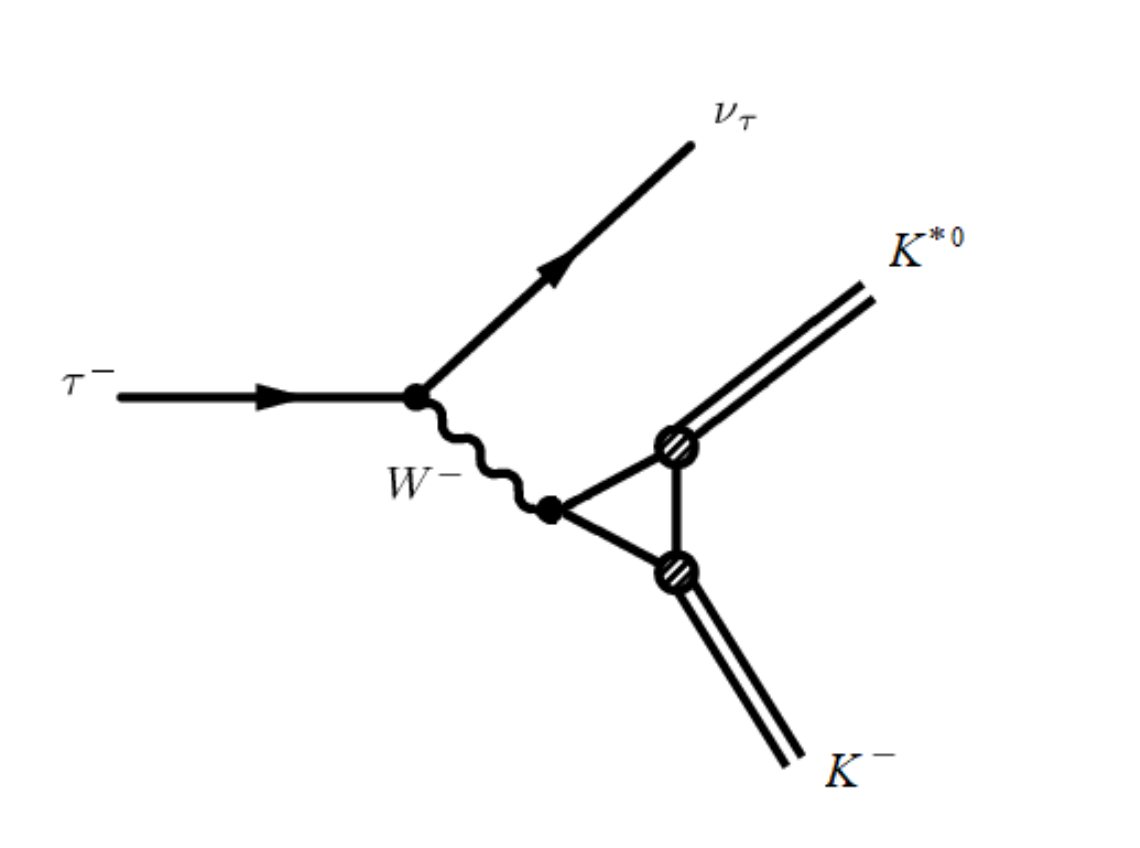}}
		\caption{Contact diagram with the production of $K^{*0}(892) K^-$ via intermediate photon.}
		\label{Contact}
	\end{figure}

\begin{figure}[h]
		\center{\includegraphics[scale = 0.8]{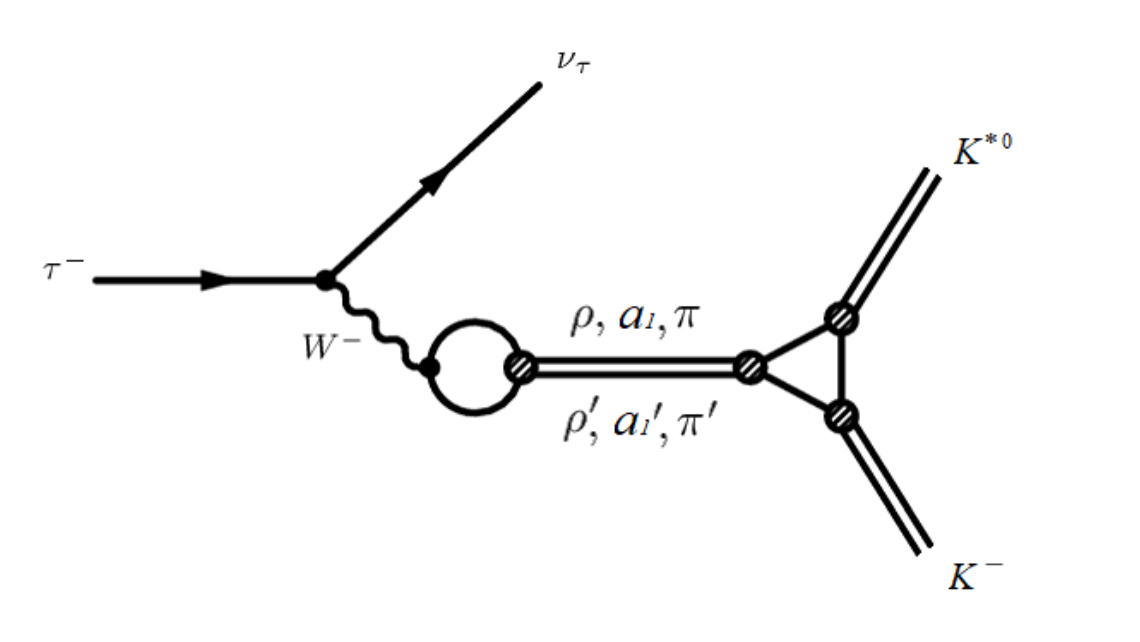}}
		\caption{Feynman diagram with the production of $K^{*0}(892) K^-$ via intermediate mesons.}
		\label{Intermediate}
	\end{figure}

	Feynman diagrams describing the process $\tau \to K^{*0}(892) K^- \nu_{\tau}$ are shown in Figs. ~\ref{Contact} and ~\ref{Intermediate}.	
	
	The corresponding amplitude in the extended NJL model is written as follows:
	\begin{eqnarray}
	\label{amplitude}
		\mathcal{M} & = & i\sqrt{2} G_{F} V_{ud} L_{\mu} \biggl[ \mathcal{M}_{c} + \mathcal{M}_{a_1} + e^{i\pi} \mathcal{M}_{a'_1} \nonumber\\ 
		&& + \mathcal{M}_{\rho} + e^{i\pi} \mathcal{M}_{\rho'} + \mathcal{M}_{\pi} + e^{i\pi}\mathcal{M}_{\pi'} \biggl]_{\mu\nu} e_{\nu}^{*}(p_{K^{*}}),
	\end{eqnarray}
where $G_{F}$ is the Fermi constant, $V_{ud}$ is the Cabibbo -- Kobayashi -- Maskawa matrix element, $L_{\mu}$ is the lepton current, and $e_{\nu}^{*}(p_{K^{*}})$ is the polarization vector of the $K^{*}(892)$ meson.  Unfortunately, the NJL model can not describe a relative phase between different states.  Thus, we take a phase from experiments \cite{Achasov:2000wy} ($e^{i \pi}$ factor in the excited mesons). 

The terms in the curly brackets (\ref{amplitude}) are the contributions of the contact diagram and diagrams with intermediate mesons in the ground and first radially excited states.	
	\begin{eqnarray}
		\mathcal{M}_{c}^{\mu\nu} & = & \left(3m_{u} - m_{s} \right)I^{K^{*}K}_{11} g^{\mu\nu} \nonumber\\
		&& - 2i\biggl[ m_{s}I^{K^{*}K}_{21} -(m_{s} - m_{u})[I^{K^{*}K}_{21} + {m^2_{u}}I^{K^{*}K}_{31}] \biggl] e^{\mu\nu\alpha\delta} p_{K\alpha} p_{K^{*}\delta}, \nonumber\\		
		\mathcal{M}_{a_1}^{\mu\nu} & = & \left(3m_{u} - m_{s}\right) \frac{C_{\rho}}{g_{\rho}}I^{a_{1}K^{*}K}_{11} \biggl[ g^{\mu\lambda} \left(q^{2} - 6 m^2_{u}\right) - q^{\mu}q^{\lambda}\biggl] BW^{\lambda\nu}_{a_{1}}, \nonumber\\
		\mathcal{M}_{a'_1}^{\mu\nu} & = & \left(3m_{u} - m_{s}\right) \frac{C_{\rho'}}{g_{\rho}}I^{a'_{1}K^{*}K}_{11} \biggl[ g^{\mu\lambda} \left(q^{2} - 6 m^2_{u}\right) - q^{\mu}q^{\lambda}\biggl] BW^{\lambda\nu}_{a'_{1}}, \nonumber\\
		\mathcal{M}_{\rho}^{\mu\nu} & = & -i2\frac{C_{\rho}}{g_{\rho}} \biggl[ m_{s}I^{\rho K^{*}K}_{12} -(m_{s} - m_{u}) [I^{\rho K^{*}K}_{12} + {m^2_{u}}I^{\rho K^{*}K}_{13}] \biggl] \nonumber\\
		&& \times \left( g^{\mu\lambda}q^{2} - q^{\mu}q^{\lambda} \right) BW_{\lambda\sigma}^{\rho} e^{\sigma\nu\alpha\delta} p_{K \alpha} p_{K^{*}\delta}, \nonumber\\
		\mathcal{M}_{\rho'}^{\mu\nu} & = & -i2\frac{C_{\rho'}}{g_{\rho}} \biggl[ m_{s}I^{\rho' K^{*}K}_{21} -(m_{s} - m_{u}) [I^{\rho' K^{*}K}_{21} + {m^2_{u}}I^{\rho' K^{*}K}_{31}] \biggl] \nonumber\\
		&& \times \left( g^{\mu\lambda}q^{2} - q^{\mu}q^{\lambda} \right) BW_{\lambda\sigma}^{\rho'} e^{\sigma\nu\alpha\delta} p_{K \alpha} p_{K^{*}\delta}, \nonumber\\
		\mathcal{M}_{\pi}^{\mu\nu} & = & -4 \biggl[ \left( \frac{m_{u} Z_{\pi}}{g_{\pi}} - \frac{6m^2_{u}}{M^2_{a_1}} \frac{C_{\rho}}{g_{\rho}}4m_{u}I^{a_1 \pi}_{20} \right)I^{\pi K^{*}K}_{11} \nonumber\\
		&& - \frac{m^2_{u} (3m_{u}-m_{s})Z_{\pi}}{M^2_{a_1} g_{\pi}} I^{a_1 \pi}_{20} I^{a_1K^{*}K}_{11} \biggl] q^{\mu}q^{\nu}BW^{\pi}, \nonumber\\
		\mathcal{M}_{\pi'}^{\mu\nu} & = & -4 \frac{m_{u}}{g_{\pi}} Z_{\pi} C_{\pi'} I^{\pi' K^{*}K}_{11} q^{\mu}q^{\nu}BW^{\pi'}.
	\end{eqnarray}
	
	Here $C_{M}$ and $C_{M'}$ are the factors arising from the quark loops describing the $W$ boson transition to an intermediate meson and are expressed as
	\begin{eqnarray}
	\label{coupling}
	C_{M} & = & \frac{1}{\sin(2\theta_{M}^{0})}\biggl[\sin(\theta_{M} + \theta_{M}^{0})	+ R_{M}\sin(\theta_{M} - \theta_{M}^{0})\biggl], \nonumber\\
	C_{M'} & = & \frac{-1}{\sin(2\theta_{M}^{0})}\biggl[\cos(\theta_{M} + \theta_{M}^{0}) + R_{M}\cos(\theta_{M} - \theta_{M}^{0})\biggl].
	\end{eqnarray}
	
The intermediate mesons are described by the Breit–Wigner propagators:	
	\begin{eqnarray}
	&BW_{M}^{\mu\nu} = \frac{g^{\mu\nu} - \frac{q^{\mu}q^{\nu}}{M_{M}^{2}}}{M_{M}^{2} - q^{2} - i \sqrt{q^{2}} \Gamma_{M}},& \nonumber\\
	& \textrm{(vector and axial-vector case)} & \nonumber\\
	& BW_{M}  =  \frac{-1}{M_{M}^{2} - q^{2} - i \sqrt{q^{2}} \Gamma_{M}}. & \\
	& \textrm{(pseudoscalar case)} & \nonumber
	\end{eqnarray}
	
	Integrals with vertices from the Lagrangian (\ref{Lagrangian}) in the numerators that also appear in the amplitude have the form
	\begin{eqnarray}
	\label{DiffIntegral}
		I_{n_{1} n_{2}}^{M, \dots, M^{'}, \dots} & = &
		-i\frac{N_{c}}{(2\pi)^{4}}\int\frac{A_{M} \dots B_{M} \dots}{(k^2 - m_{u}^{2})^{n_{1}}(k^2 - m_{s}^{2})^{n_{2}}} \Theta(\Lambda^{2} - {\bf k}^2)	\mathrm{d}^{4}k,
	\end{eqnarray}
	where $A_{M}$ and $B_{M}$ are given in (\ref{verteces}).
	
	The results obtained by using this amplitude are shown in Table 1 and the prediction for the differential distribution over the invariant mass of the meson pair $K^{*0}(892) K^-$ is presented in Fig. 3. 
		
	\begin{table}[h!]
	\label{Tabddd}
	\begin{center}
	\begin{tabular}{cc} \hline
	Channel & Br $(\times 10^{-3})$  \\ \hline
	AV & $ 1.01 $  \\
	AV' & $ 0.000018 $  \\
	V & $ 0.32 $  \\
	V' & $ 0.24 $  \\
	PS & $ 0.09 $  \\
	PS' & $ 0.000035 $  \\ \hline
	Total & $ 1.99$  \\ \hline
	\end{tabular}
	\end{center}
	\caption{Separate contributions to the process $\tau\to K^{*0}(892) K^- \nu_{\tau}$ in the extended NJL model.}
	\end{table}
	
	\begin{figure}[h]
	\label{width}
	\center{\includegraphics[scale = 1.0]{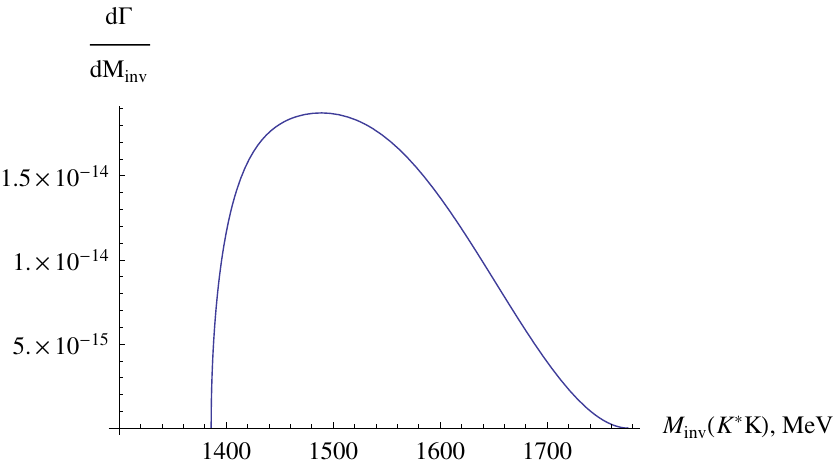}}
	\caption{ The differential decay width for the process $\tau \to K^{*0}(892) K^- \nu_{\tau}$.}
	\label{Width}
	\end{figure}
	
	The experimental value for the decay width of this process \cite{Tanabashi:2018oca} is 
	\begin{eqnarray}
		Br(\tau \to K^{*0}(892) K^- \nu_{\tau})_{exp} = (2.1 \pm 0.4) \times 10^{-3}.
	\end{eqnarray} 
	
\section{Conclusion}
	In the present work, a satisfactory description of the decay $\tau \to K^{*0}(892) K^- \nu_{\tau}$ was obtained in the framework of the extended NJL model. 
	Note that this decay differs from other similar $\tau$ decays with a pair of vector and pseudoscalar strange mesons in the final state by the following features: first, this process includes the contributions of intermediate channels with non-strange $a_1$, $\rho$ and $\pi$ mesons. Second, for the first time in $\tau$ decays with a pair of vector and pseudoscalar mesons in the final state with strange mesons, the dominant role is played not only by the axial-vector channel but also by the vector channel. Third, in previous $\tau$ decays, intermediate channels with radially excited mesons give a negligible contribution, while in the process under consideration the intermediate vector channel with the $\rho'$ meson gives a significant contribution. These qualities make us pay attention to the process $\tau \to K^{*0}(892) K^- \nu_{\tau}$ considered here. As a result, satisfactory agreement with the experimental data were obtained without any additional arbitrary free parameters. Also, a prediction for the differential decay width $\tau \to K^{*0}(892) K^- \nu_{\tau}$ is presented in Fig. 3. 
	
	In \cite{Li:1996md}, the calculations were performed within the framework of the chiral-symmetric model, where intermediate mesons were considered only in the ground state. As a result, the value for the partial width $Br(\tau \to K^{*0}(892) K^- \nu_{\tau}) = 3.92 \times 10^{-3}$ was obtained. The main contribution to the decay width comes from only vector channel with the $\rho$ meson. Also, the theoretical result for the width $Br(\tau \to K^{*0}(892) K^- \nu_{\tau}) = 4.93 \times 10^{-3}$, exceeding the experimental value, was obtained in the paper \cite{Dai:2018thd}. In \cite{Guo:2008sh}, the theoretical value $Br(\tau \to K^{*0}(892) K^- \nu_{\tau}) = 1.5 \times 10^{-3}$ was obtained by using the chiral theory with resonances . 
	
\section*{Acknowlegments}
The authors are grateful to A. B. Arbuzov for useful discussions; this work is supported by the JINR grant for young scientists and specialists No.19-302-06.

\end{document}